\documentclass[a4paper,11pt]{article}

\usepackage{pos}
\usepackage{subcaption}
\usepackage[backend=biber,style=numeric-comp,sorting=none]{biblatex}
\newcommand{\Mtree}{\mathcal{M}_{ij\to ij}^{\text{tree}}}
\newcommand{\T}{\mathbf{T}}
\newcommand{\eps}{\epsilon}
\newcommand{\M}{\mathcal{M}}
\newcommand{\al}{\alpha}
\newcommand{\nn}{\nonumber}

\renewcommand{\logo}{\relax}

\title{High-energy limit of $2\to 2$ scattering amplitudes at NNLL}

\author*[a]{Calum Milloy}
\author[b]{Giulio Falcioni}
\author[b]{Einan Gardi}
\author[b]{Niamh Maher}
\author[a,c]{Leonardo Vernazza}

\affiliation[a]{Dipartimento di Fisica and Arnold-Regge Center,\\
 Universit\`{a} di Torino, and INFN, Sezione di Torino, Via P. Giuria 1, I-10125 Torino, Italy}

\affiliation[b]{Higgs Centre for Theoretical Physics, School of Physics and Astronomy,\\
 The University of Edinburgh, Edinburgh EH9 3FD, Scotland, UK}

\affiliation[c]{Theoretical Physics Department, CERN,
 Geneva 1211, Switzerland}

\emailAdd{calumwilliam.milloy@unito.it}

\abstract{The high-energy limit of $2\to 2$ scattering amplitudes offers an excellent setting to explore the universal features of gauge theories. At Leading Logarithmic (LL) accuracy the partonic amplitude is governed by Regge poles in the complex angular momentum plane. Beyond LL, Regge cuts in this plane begin to play an important role. Specifically, the real part of the amplitude at Next-to-Next-to-Leading Logarithmic (NNLL) accuracy presents for the first time both a Regge pole and a Regge cut. Analysing this tower of logarithms and computing it explicitly through four loops we are able to systematically separate between the Regge pole and the Regge cut. The former involves two fundamental parameters, namely the gluon Regge trajectory and impact factors. We explain how to consistently define the impact factors at two loops and the Regge trajectory at three loops. We confirm that the singularities of the trajectory are given by the cusp anomalous dimension. We also show that the Regge-cut contribution at four loop is nonplanar.}

\FullConference{%
  Loops and Legs in Quantum Field Theory - LL2022,\\
  25-30 April, 2022\\
  Ettal, Germany
}

\begin{document}
\maketitle

\section{Introduction to the high-energy limit of $2\to 2$ scattering amplitudes}

The high-energy limit has long been an intersting laboratory to study gauge-theory amplitudes, which simplify significantly in this regime. The simplification in the limit means that multi-loop corrections are accessible, and they may even be resummed to all orders (at a fixed logarithmic accuracy), while still retaining a rich structure in both colour and kinematics.

\begin{figure}[h]
  \centering
  \includegraphics[width=0.35\textwidth]{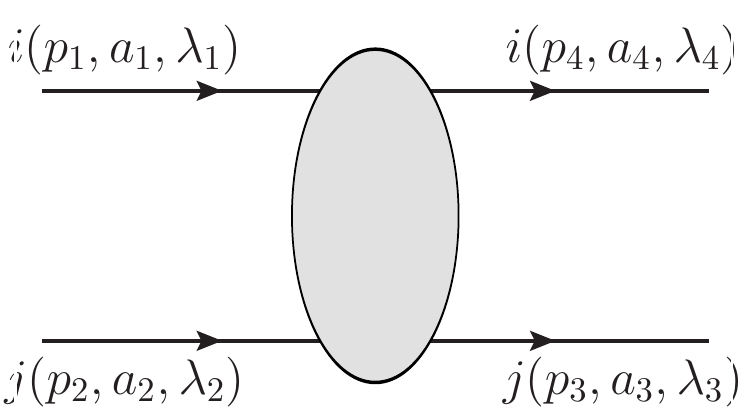}
  \caption{A representation of the $2\to 2$ scattering of eq.~(\ref{eq:22Scatter}).}
  \label{fig:figure1}
\end{figure}

We consider the $2\to 2$ scattering of massless partons in a generic gauge theory
\begin{align}\label{eq:22Scatter}
  i(p_1,a_1,\lambda_1)
  +
  j(p_2,a_2,\lambda_2)
  \to
  j(p_3,a_3,\lambda_3)
  +
  i(p_4,a_4,\lambda_4),
\end{align}
where momenta, colour and helicity are labelled by $p_i$, $a_i$ and $\lambda_i$ respectively. A pictorial representation is given in Figure~\ref{fig:figure1}. The process is described in terms of the Mandelstam invariants,
\begin{align}
  s = (p_1+p_2)^2 > 0
  &&
  t = (p_1-p_4)^2 < 0
  &&
  u = (p_1-p_3)^2 < 0,
\end{align}
where the signs dictate the physical scattering region. Momentum conservation implies
\begin{align}\label{eq:momCons}
  s + t + u = 0.
\end{align}
In $2\to 2$ scattering, the high-energy (Regge) limit is defined to be when the centre-of-mass energy is much larger than the momentum transfer, i.e. $s\gg -t$. At leading power the tree-level amplitude is
\begin{align}\label{eq:Mtree}
  \Mtree
  &=
  4\pi\al_s\frac{2s}{t}
  (\T_i^b)_{a_1a_4}(\T_j^b)_{a_2a_3}\delta_{\lambda_1\lambda_4}\delta_{\lambda_2\lambda_3},
\end{align}
where we observe that the helicity of each of the scattered partons is conserved. We use colour operator notation to keep expressions general for any representation. In the Regge limit it is convenient to define colour-flow operators following the conventions of the Mandelstams~\cite{Dokshitzer:2005ig,DelDuca:2011wkl,DelDuca:2011ae},
\begin{align}
  \T_s = \T_1 + \T_2,
  &&
  \T_t = \T_1 + \T_4,
  &&
  \T_u = \T_1 + \T_3.
\end{align}
The quantity $\T_t^2$ measures the colour charge of the $t$-channel adjoint exchange,
\begin{align}
\T_t^2\Mtree = C_A\Mtree.
\end{align}

In the high-energy limit it can be shown that in the complex angular momentum plane, the amplitude can be written as a sum over poles and cuts, dubbed \emph{Regge poles} and \emph{Regge cuts}. Regge cuts were shown by Mandelstam to arise from only nonplanar diagrams \cite{Mandelstam:1963cw}. Note that Mandelstam's observation only involved the \emph{diagrams} being nonplanar. In this work we connect this to gauge theories, where planarity relates to colour and corresponds to the large-$N_c$ limit.

Calculating the amplitude in the perturbative expansion in $\al_s$ we would find that large logarithms will develop in the ratio $s/(-t)$. In the Leading Logarithmic (LL) approximation, these can be resummed by the famous Regge pole,
\begin{align}\label{eq:LL}
  \M_{ij\to ij}^{\text{LL}} = \left(\frac{s}{-t}\right)^{C_A\al_g(t,\mu^2)}\Mtree,
\end{align}
where $\al_g$ is called the Regge trajectory. This quantity admits a perturbative expansion,
\begin{align}\label{eq:alExpansion}\al_g(t,\mu^2)=\sum_{n=1}^\infty\left(\frac{\al_s}{\pi}\right)^n\al_g^{(n)}(t,\mu^2),\end{align}
where the one-loop coefficient relevant at LL is
\begin{align}
  \al_g^{(1)}(t,\mu^2) = \frac{r_\Gamma}{2\eps}\left(\frac{\mu^2}{-t}\right)^\eps,
  &&
  r_\Gamma = e^{\eps\gamma_E}\frac{\Gamma^2(1-\eps)\Gamma(1+\eps)}{\Gamma(1-2\eps)},
\end{align}
with $\eps=(4-d)/2$ the dimensional regularisation parameter. We will set the renormalisation scale $\mu^2=-t$ for simplicity. At this accuracy the amplitude has the same colour structure as the tree-level, the octet exchange in the $t$-channel.

Before moving on to Next-to-Leading Logarithms (NLL), we shall first introduce the concept of signature under $s\leftrightarrow u$, where we notice that the tree-level amplitude in eq.~(\ref{eq:Mtree}) is signature-odd. We define even and odd amplitudes through
\begin{align}
  \M^{(\pm)}(s,t) = \frac12\left(\M(s,t) \pm \M(-s-t,t)\right).
\end{align}
Upon defining the signature-even logarithm,
\begin{align}
  L \equiv \log\left(\frac{s}{-t}\right) - \frac{i\pi}{2} 
  = 
  \frac12\left[\log\left(\frac{-s-i0}{-t}\right)
  +
  \log\left(\frac{-u-i0}{-t}\right)\right],
\end{align}
and then expanding the amplitude in $L$, the coefficients of $L$ in the odd (even) amplitudes are entirely real (imaginary)~\cite{Caron-Huot:2017fxr}. 

In this work, we focus solely on the odd amplitude $\mathcal{M}^{(-)}$, and we will remove the superscript $(-)$ for convenience. For even amplitudes at NLL see~\cite{Caron-Huot:2013fea,Caron-Huot:2017zfo,Caron-Huot:2020grv}. At NLL, the amplitudes obey Regge pole factorisation~\cite{Fadin:2006bj,Ioffe:2010zz,Fadin:2015zea},
\begin{align}\label{eq:NLLFact}
  \M_{ij\to ij}^{\text{NLL}} = e^{C_A\al_g(t)L}C_i(t)C_j(t)\Mtree,
\end{align}
where $C_i(t)$ are energy-independent impact factors for each of the scattered partons. They are expanded in powers of $\al_s$, similar to $\al_g$ in eq.~(\ref{eq:alExpansion}). The dependence on $s$ is entirely controlled by the Regge trajectory. NLL accuracy is obtained by taking the Regge trajectory up to two loops, and the impact factors at one loop.

It is worthwhile to compare Regge factorisation to infrared factorisation. It is well-known that long-distance singularities factorise from the amplitude~\cite{Catani:1998bh,Sterman:2002qn,Aybat:2006mz,Aybat:2006wq,Ma:2019hjq,Feige:2014wja},
\begin{align}\label{eq:IRFact}
  \M_{ij\to ij} = \mathbf{P}\exp\left\{-\frac12\int_0^{\mu^2}\frac{d\lambda^2}{\lambda^2}\mathbf{\Gamma}_{ij\to ij}\right\}\cdot\mathcal{H}
\end{align}
where $\mathbf{\Gamma}_{ij\to ij}$ is dubbed the soft anomalous dimension. In the high-energy limit it is given by~\cite{Caron-Huot:2017fxr}
\begin{align}\label{eq:softAD}
  \mathbf{\Gamma}_{ij\to ij} = \frac12\gamma_K\left[L\T_t^2+i\pi\T_{s-u}^2\right]+\Gamma_i+\Gamma_j+\mathbf{\Delta}
\end{align}
where at LL, divergences are captured by the lightlike cusp anomalous dimension $\gamma_K$ and $\Gamma_i$ is the collinear anomalous dimension, which collects collinear divergences associated to scattered parton~$i$. The operator $\T_{s-u}^2$ is defined as
\begin{align}
  \T_{s-u}^2 = \frac12\left(\T_s^2-\T_u^2\right).
\end{align}
The quantity $\mathbf{\Delta}$ starts at three loops~\cite{Gardi:2019pmk} but only contributes at NLL for even amplitudes from four loops~\cite{Caron-Huot:2013fea} and NNLL for odd amplitudes also from four loops~\cite{Falcioni:2020lvv}. Performing the integral over the $d$-dimensional running coupling in eq.~(\ref{eq:IRFact}), reveals the explicit infrared poles of the amplitude.

It is natural to compare the two exponentiations in eqs.~(\ref{eq:NLLFact}) and~(\ref{eq:IRFact}). The coefficient of the logarithm $L$ in the exponent gives
\begin{align}\label{eq:reggeCusp}
   \al_g(t) \stackrel{?}{=} -\frac14\int_0^{\mu^2}\frac{d\lambda^2}{\lambda^2}\gamma_K(\al_s(\lambda^2)) + \mathcal{O}(\eps^0).
 \end{align} 
The equivalency holds at two loops (i.e. NLL)~\cite{Korchemskaya:1996je}. We shall show that it also holds at three loops through a proper definition of the trajectory and it remains a conjecture that it holds beyond.
The factor independent of the logarithm suggests that the infrared divergences of the impact factors are related to the collinear anomalous dimension $\Gamma_i$~\cite{Falcioni:2021buo,DelDuca:2014cya}.

\section{Distinguishing pole and cut contributions}

At NNLL the factorisation in eq.~(\ref{eq:NLLFact}) fails. The colour structure of the amplitude is no longer given only by the exchange of an octet in the $t$-channel, as for the tree-level amplitude, and new colour channels open up. To describe also these contributions to the amplitudes, we generalise eq.~(\ref{eq:NLLFact}) by introducing a non-factorising term, which originates from Regge cuts~\cite{DelDuca:2001gu,DelDuca:2013ara,DelDuca:2013dsa,DelDuca:2014cya,Caron-Huot:2017fxr,Fadin:2016wso,Fadin:2017nka,Fadin:2020lam,Fadin:2021csi,Falcioni:2020lvv,Falcioni:2021buo}
\begin{align}\label{eq:NNLL}
  \M_{ij\to ij}^{\text{NNLL}} = e^{C_A\al_g(t)L}C_i(t)C_j(t)\Mtree + \M_{ij\to ij}^{\text{NF}}.
\end{align}

The high-energy behaviour of the amplitude will only be uncovered when the perturbative series is resummed. As such, it is not clear how to properly define the Regge pole and Regge cut in order-by-order computations. To see this explicitly, we expand eq.~(\ref{eq:NNLL}) to two loops
\begin{align}\label{eq:NNLL2Loop}
  \mathcal{M}_{ij\to ij}^{(2),\text{NNLL}}
  =
  \left(
  C_i^{(1)}C_j^{(1)}
  + C_i^{(2)}
  + C_j^{(2)}
  \right)
  \Mtree
  +
  \M_{ij\to ij}^{(2),\text{NF}}.
\end{align}
In the expansion above, we are a priori free to move any term whose colour factor is proportional to the tree-level from $\M_{ij\to ij}^{(2),\text{NF}}$ into the two-loop impact factors $C_i^{(2)}$ (the one-loop $C_i^{(1)}$ are fixed at NLL). There is an ambiguity in the definition of the two-loop impact factors. Therefore, we need a prescription to disentangle the two-loop impact factors from the contribution of Regge cuts. How to properly distinguish the pole and cut contributions is the key result in what follows~\cite{Falcioni:2021dgr}.

We can access this non-factorising term by computing certain diagrams in the high-energy limit. In this regime, the fundamental degrees of freedom are Reggeized gluons, dubbed \emph{Reggeons}, see~\cite{Caron-Huot:2013fea,Fadin:2020lam} and references therein. At tree-level, the diagram featuring the exchange of a single Reggeon in the $t$-channel has the analytic structure of a Regge pole, i.e. the first term in eq.~(\ref{eq:NNLL}).

\begin{figure}
  \centering
  \begin{subfigure}[t]{0.3\textwidth}
    \centering
    \includegraphics[width=0.8\textwidth,height=0.55\textwidth]{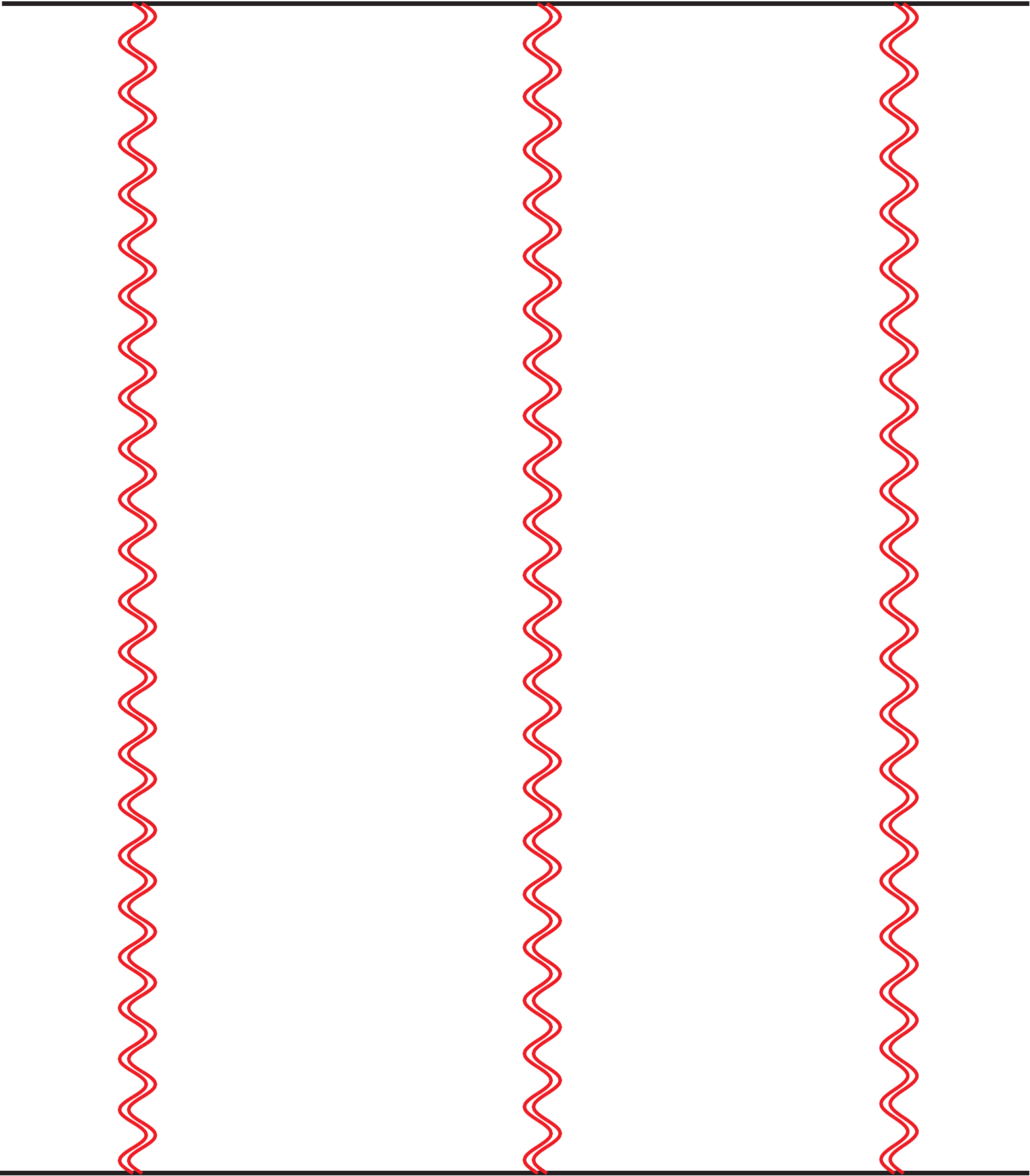}
    \caption{A two-loop MR diagram. The explicit result is given in eq.~(\ref{eq:MRSTwoLoop}).}
    \label{fig:figure2}
  \end{subfigure}
  \hfill
  \begin{subfigure}[t]{0.3\textwidth}
    \centering
    \includegraphics[width=0.8\textwidth,height=0.55\textwidth]{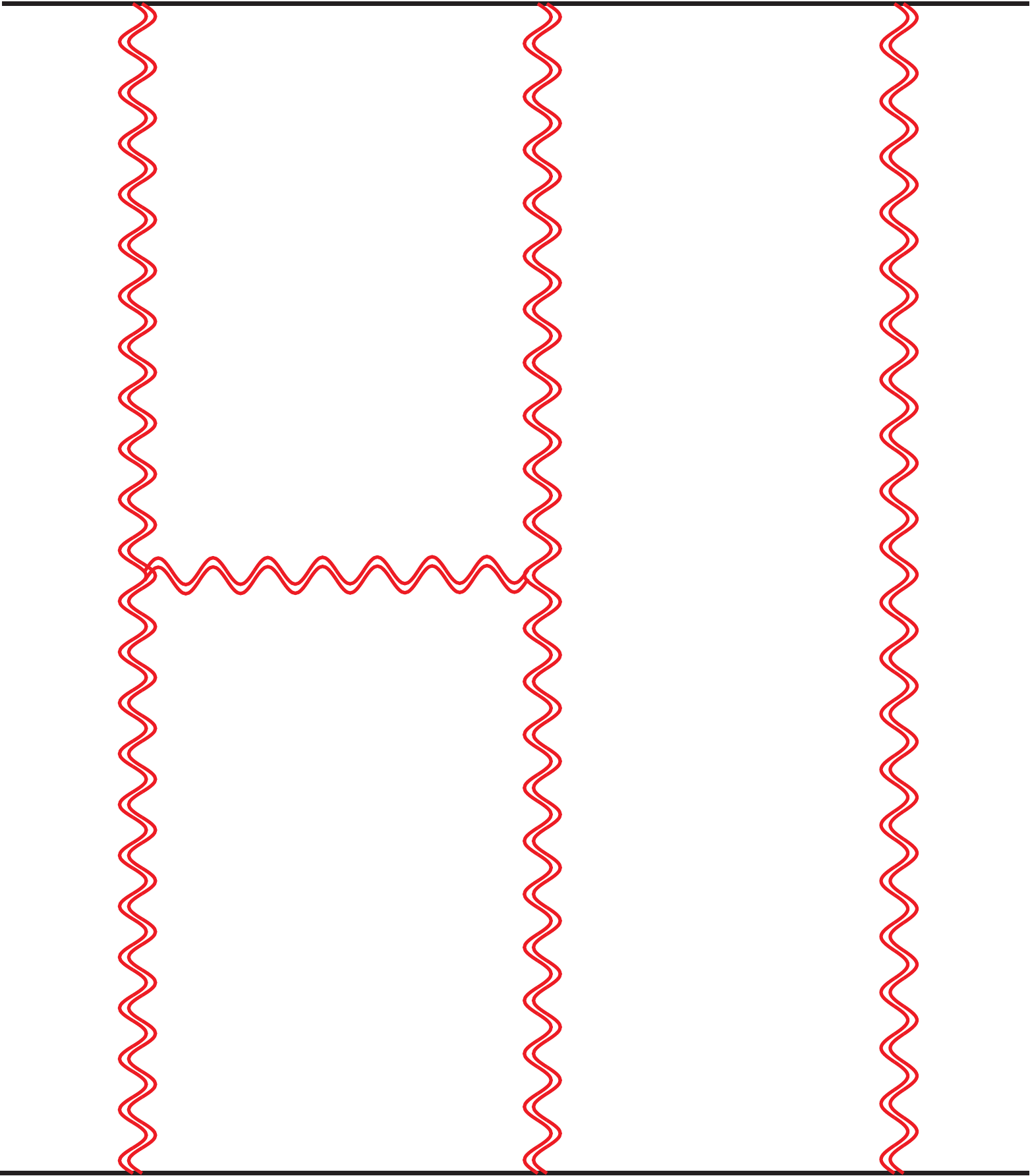}
    \caption{A three-loop MR diagram.}
    \label{fig:figure3}
  \end{subfigure}
  \hfill
  \begin{subfigure}[t]{0.3\textwidth}
    \centering
    \includegraphics[width=0.8\textwidth,height=0.55\textwidth]{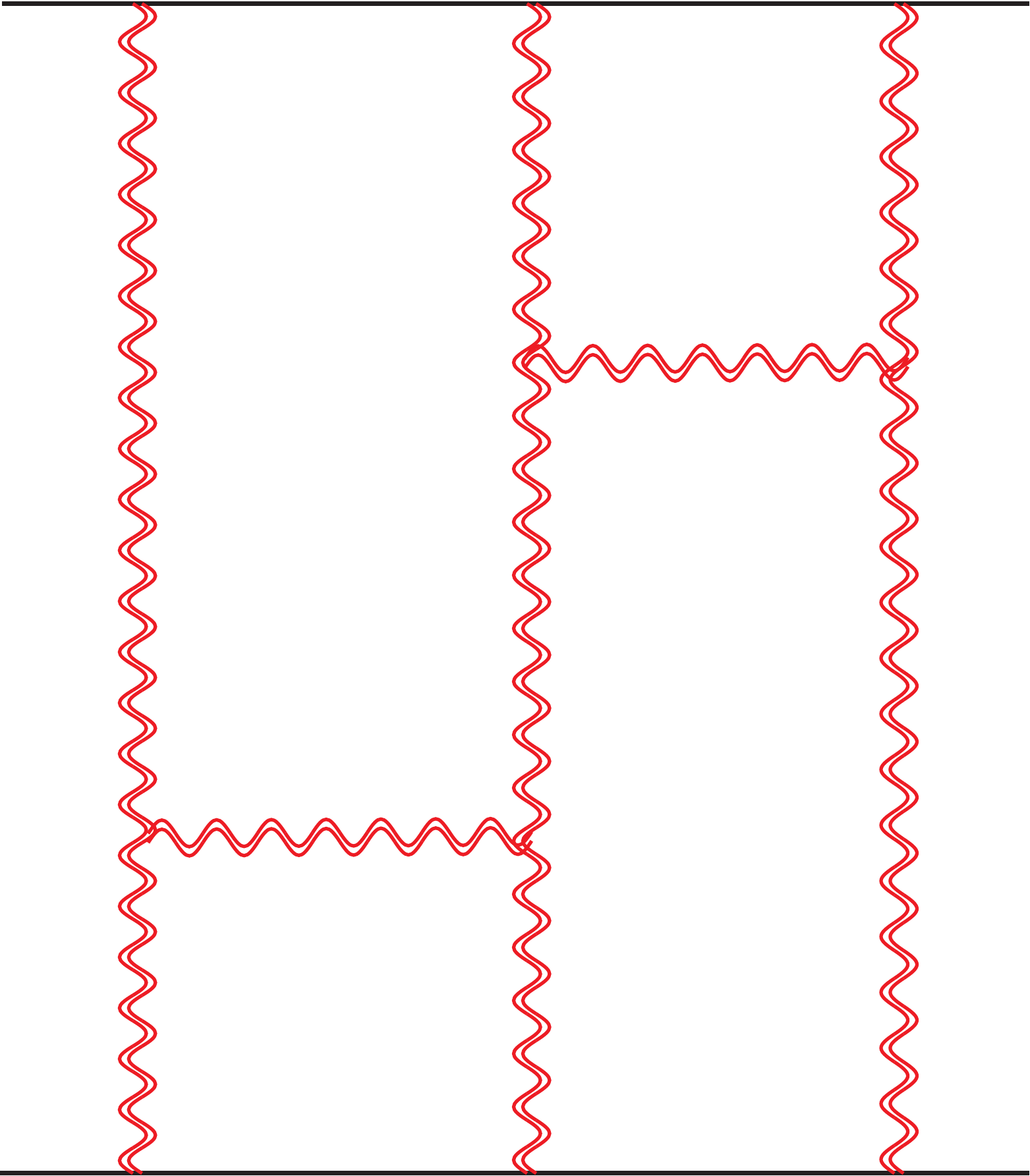}
    \caption{A four-loop MR diagram.}
    \label{fig:figure4}
  \end{subfigure}
  \caption{Example MR diagrams. The black lines represent the external scattering partons and the double red lines are the Reggeon exchanges. There is a symmetric sum over all possible orderings of the Reggeon attachments to the external partons.}\label{fig:diagrams}
\end{figure}

We use the shockwave formalism of ref.~\cite{Caron-Huot:2013fea} to compute diagrams with the exchange of Multiple Reggeons (MR), which start at two loops\footnote{At one-loop, it is possible to exchange only two Reggeons, but these contribute only to the imaginary (even) part of the amplitude.}. Example diagrams at two, three and four loops are given in Figure~\ref{fig:diagrams}. The two-loop diagram displayed in Figure~\ref{fig:figure2} was computed in~\cite{Caron-Huot:2017fxr}, giving
\begin{align}\label{eq:MRSTwoLoop}
  \mathcal{M}_{ij\to ij}^{(2),\text{MR}}
  =
  \pi^2
  \left(
  -\frac1{8\eps^2}
  +\frac18\zeta_2
  +\frac43\eps\zeta_3
  +\mathcal{O}(\eps^2)
  \right)
  \left(
  (\T_{s-u}^2)^2-\frac1{12}C_A^2
  \right)
  \Mtree.
\end{align}
A couple of comments are in order. The overall $\pi^2=-(i\pi)^2$ highly suggests it originates from a cut. It has uniform transcendental weight as this quantity is independent on the underlying gauge theory: the $\mathcal{N}=4$ super Yang-Mills (SYM) result is the same as QCD.

These features are maintained in the MR exchanges at three and four loops, which are computed in~\cite{Caron-Huot:2017fxr} and~\cite{Falcioni:2020lvv}, respectively. It is possible to identify the entire contribution of MR diagrams with the Regge-cut term in eq.~(\ref{eq:NNLL}), as done in~\cite{Caron-Huot:2017fxr}. This defines a scheme to extract the two-loop impact factors $C_i^{(2)}$ and the three-loop Regge trajectory $\al_g^{(3)}$ by matching eq.~(\ref{eq:NNLL}), where $\M_{ij\to ij}^{\text{NF}}$ is identified with $\mathcal{M}_{ij\to ij}^{\text{MR}}$, with fixed-order calculations~\cite{Ahmed:2019qtg,Henn:2016jdu,Caola:2021rqz,Caola:2021izf,Caola:2022dfa}.

Here, we take a different point of view following refs.~\cite{Falcioni:2021buo,Falcioni:2021dgr}. Indeed, according to the analysis of Mandelstam~\cite{Mandelstam:1963cw}, the Regge cut contribution $\M_{ij\to ij}^{\text{NF}}$ arises from nonplanar diagrams. Therefore, its colour structure should be subleading in the limit of large $N_c$. In contrast with this conclusion, the $\mathcal{M}_{ij\to ij}^{\text{MR}}$ has non-vanishing contributions at two and three loops for large $N_c$~\cite{Falcioni:2021buo,Falcioni:2020lvv},
\begin{align}\label{eq:MRSPlanar}
  \mathcal{M}_{ij\to ij}^{\text{MR}}\rvert_{\text{planar}}
  =&
  \frac{\pi^2r_\Gamma^2N_c^2}{6}
  \left(\frac{\al_s}{\pi}\right)^2
  \left\{
  S^{(2)}(\eps)
  -
  \left(\frac{\al_s}{\pi}\right)
  r_\Gamma N_c L
  \left[
  S^{(3)}(\eps)
  -\frac1{2\eps}S^{(2)}(\eps)
  \right]
  +\mathcal{O}(\al_s^3)
  \right\}
  \Mtree,
  \\
  S^{(2)}(\epsilon)  =& -\frac{1}{8\epsilon^2}+\frac{3}{4}\epsilon\zeta_3+\frac{9}{8}\epsilon^2\zeta_4+{\cal{O}}(\epsilon^3),
  \\
  S^{(3)}(\eps) =& - \frac1{144\eps^3}+\frac{35}{72}\zeta_3+\frac{35}{48}\eps\zeta_4+{\cal{O}}(\epsilon^3).
\end{align}
By looking at eq.~(\ref{eq:MRSPlanar}), we notice that, for every scattering process, the planar part of $\mathcal{M}_{ij\to ij}^{\text{MR}}$ is colour proportional to the tree-level amplitude. Because of this property, by following the argument below eq.~(\ref{eq:NNLL2Loop}), at two loops it is possible to absorb the planar part of $\mathcal{M}_{ij\to ij}^{\text{MR}}$ in the definition of the impact factors~\cite{Falcioni:2021buo,Falcioni:2021dgr}. At three loops, the planar terms in $\mathcal{M}_{ij\to ij}^{\text{MR}}$ are again proportional to the tree-level amplitude and can be absorbed in the definition of the three-loop gluon Regge trajectory. This naturally gives a method to capture the true separation between the pole and the cut, where we shift all the planar corrections into a new pole term with new impact factors and a new Regge trajectory~\cite{Falcioni:2021dgr}
\begin{align}\label{eq:tildePoleTerm}
  \M_{ij\to ij}^{\text{pole}}
  =
  \tilde C_i(t)
  \tilde C_j(t)
  e^{\tilde\al_g(t)C_AL}
  \Mtree,
\end{align}
where we have added a tilde on the parameters to dinstinguish between the old and new pole terms. As the whole amplitude is a well-defined quantity, the shifts in the two-loop impact factors and three-loop Regge trajectory can be calculated
\begin{align}
  \label{eq:ImpactFactorShift}
  \tilde C_{i/j}^{(2)}
  =&\,
  C_{i/j}^{(2)}
  +
  N_c^2\,(r_\Gamma)^2\frac{\pi^2}{12}S^{(2)}(\eps),
  \\
  \label{eq:ReggeTrajShift}
  \tilde \al_g^{(3)}
  =&\,
  \al_g^{(3)}
  -N_c^2(r_\Gamma)^3\frac{\pi^2}{6}S^{(3)}(\eps).
\end{align}
This exhausts all possible parameters at NNLL to absorb planar contributions into the pole term. This is consistent with the fact that at four loops the MR term is entirely nonplanar~\cite{Falcioni:2020lvv,Falcioni:2021buo} and that it is conjectured to be so at all higher loops~\cite{Falcioni:2021buo,Falcioni:2021dgr}.

\section{Applications}

Having properly distinguished pole and cut contributions order-by-order in the perturbation expansion, we can now extract Regge parameters. Using state-of-the-art calculations for the relevant amplitudes, we first extract two-loop impact factors and then the three-loop Regge trajectory. Lastly, as a by-product of the four-loop computation of $\mathcal{M}_{ij\to ij}^{(4),\text{MR}}$ we derive constraints on the infrared structure of four-loop amplitudes.

\subsection{Two-loop impact factors}
Using eq.~(\ref{eq:ImpactFactorShift}) we can extract the two-loop impact factors. To display results it is more convenient to define the pole term as a symmetric sum of exponentials, rather than a single exponentiated sum. In eq.~(\ref{eq:tildePoleTerm}) we make the replacement
\begin{align}
  \tilde C_i(t)
  \tilde C_j(t)
  e^{\tilde\al_g(t)C_AL}
  \to
  Z_i(t)
  \bar D_i(t)
  \left[
  \left(\frac{s}{-t}\right)^{\tilde\al_g(t)}
  +
  \left(\frac{-s}{-t}\right)^{\tilde\al_g(t)}
  \right]
  Z_j(t)
  \bar D_j(t),
\end{align}
where $Z_i(t)$ is the exponential of the integral over the collinear anomalous dimension $\Gamma_i$
\begin{align}
   Z_i(t) \equiv&\exp\left\{-\frac12\int_0^{\mu^2}\frac{d\lambda^2}{\lambda^2}\Gamma_i\right\}.
 \end{align} 
This replacement defines a finite collinear-subtracted impact factor $\bar D_i(t)$, i.e.~the poles of the impact factor are given by the collinear anomalous dimension~\cite{DelDuca:2014cya}. It is related to the $\tilde C_i(t)$ by
\begin{align}
  \tilde C_i(t)
  =
  Z_i(t)
  \bar D_i(t)
  \sqrt{
  \cos{}
  \left(\frac{\pi C_A\tilde\al_g(t)}{2}\right)
  }.
\end{align}

Upon matching to amplitudes calculated in~\cite{Ahmed:2019qtg} we find the results
\begin{align}
  \bar{D}_g^{(2)}=&\, C_A^2\Big(\frac{335}{288}\zeta_2+\frac{11}{18}\zeta_3-\frac{3}{32}\zeta_4-\frac{26675}{10368}\Big)
    +C_A n_f\Big(\frac{49}{108}-\frac{25}{144}\zeta_2+\frac{5}{36}\zeta_3\Big)
    \nonumber\\&
    +C_F n_f\!\Big(\frac{55}{192}-\frac{\zeta_3}{4}\Big)
    -\frac{25}{2592}n_f^2+ {\cal O}(\epsilon)\,,
    \label{eq:gluonDg}
    \\
\bar{D}_q^{(2)}=&\, \nn
 C_A^2 \bigg(\frac{73 \zeta_2}{32} 
- \frac{43 \zeta_3}{48} 
- \frac{53 \zeta_4}{64} 
+\frac{13195}{3456} \bigg)
+C_A C_F \bigg(  
- \frac{5 \zeta_2}{2} 
+ \frac{475 \zeta_3}{144} 
+ \frac{65 \zeta_4}{32} 
- \frac{78229}{10368} \bigg)
\nonumber\\& 
+ C_F^2 \bigg(
\frac{21 \zeta_2}{16} 
- \frac{15 \zeta_3}{8}
- \frac{83 \zeta_4}{64} 
+\frac{255}{128} \bigg) 
+ C_A n_f \bigg(
- \frac{5\zeta_2}{16} 
- \frac{7 \zeta_3}{24}
-\frac{385}{432} \bigg),
\nonumber\\&
+ C_F n_f \bigg(
 \frac{\zeta_2}{8} 
+\frac{19 \zeta_3}{72}
+\frac{505}{648} \bigg) 
+ \frac{25}{864}n_f^2  + {\cal O}(\epsilon)\,
\label{eq:quarkDq}
\end{align}
for gluons and quarks respectively. The values through to ${\cal O}(\epsilon^2)$ are given in ref.~\cite{Falcioni:2021dgr}. Upon taking the planar limit we find that $\bar{D}_g^{(2)}$ agrees with the recent calculation of ref.~\cite{DelDuca:2021vjq}. Upon extracting the terms of highest weight of eq.~(\ref{eq:gluonDg}), we find the gluon impact factor in planar $\mathcal{N}=4$ SYM~\cite{DelDuca:2008pj,DelDuca:2008jg}, which also coincides with the impact factor in full colour~\cite{Henn:2016jdu,Caron-Huot:2017fxr} when using the cut scheme.

\subsection{Three-loop Regge trajectory}

We can also extract the Regge trajectory. We use eq.~(\ref{eq:ReggeTrajShift}) and the state-of-the-art three-loop four quark amplitudes of ref.~\cite{Caola:2021rqz}. In doing so, we find the poles are given by the cusp anomalous dimension, thus confirming eq.~(\ref{eq:reggeCusp}) at three loops. If we had used the MR scheme, we would find that there are infrared poles not described by $\gamma_K$~\cite{Caron-Huot:2017fxr}.

Upon defining the cusp-subtracted Regge trajectory $\hat{\tilde\al}_g$ through
\begin{align}\label{eq:cuspSubtractedTraj}
  \hat{\tilde\al}_g \equiv
  \tilde\al_g
  +
  \frac14
  \int_0^{\mu^2}\frac{d\lambda^2}{\lambda^2}\gamma_K(\al_s(\lambda^2))
\end{align}
we find the finite result at three loops~\cite{Falcioni:2021dgr,Caola:2021izf}
\begin{align}
\label{alpha_g_3}
\hat{\tilde\al}_g^{(3)} =
&\, C_A^2 \bigg(\frac{297029}{93312}-\frac{799 \zeta_2}{1296}-\frac{833 \zeta_3}{216}-\frac{77 \zeta_4}{192}+\frac{5}{24} \zeta_2 \zeta_3+\frac{\zeta_5}{4}\bigg)
\nn
\\&
+C_A n_f \bigg(\frac{103 \zeta_2}{1296}+\frac{139 \zeta_3}{144}-\frac{5 \zeta_4}{96}-\frac{31313}{46656}\bigg)
 +C_F n_f \bigg(\frac{19 \zeta_3}{72}+\frac{\zeta_4}{8}-\frac{1711}{3456}\bigg)
 \nn
 \\
 &
  +n_f^2 \left(\frac{29}{1458}-\frac{2\zeta_3}{27}\right) + \mathcal{O}(\eps).
\end{align}
Notice that when $n_f=0$ there are no $N_c$-subleading terms, in agreement with the expectation that the Regge trajectory is maximally non-Abelian~\cite{Korchemskaya:1996je}. The result of eq.~(\ref{alpha_g_3}) agrees with the planar extraction given in~\cite{DelDuca:2021vjq}.

\subsection{Infrared Constraints} 
\label{sub:infrared_constraints}

Another application of the calculation of the MR term is that we can find constraints on the infrared structure of gauge-theory amplitudes. The structure of infrared divergences for multi-leg massless scattering is known at three loops in general kinematics~\cite{Almelid:2015jia}. In addition, in the high-energy limit the entire NLL tower of corrections has been resummed~\cite{Caron-Huot:2017zfo,Gardi:2019pmk}. Given that the four-loop computation of the Regge-cut term is known, we can match the Regge decomposition in eq.~(\ref{eq:NNLL}) to the soft anomalous dimension in the high-energy limit, eq.~(\ref{eq:softAD}). It can be shown that only the Regge-cut term contributes to $\mathbf{\Delta}$. Expanding $\mathbf{\Delta}$ as
\begin{align}
  \mathbf{\Delta} = \sum_{m=3}^\infty\left(\frac{\al_s}{\pi}\right)^m \sum_{n=0}^{m-1}
  L^n\mathbf{\Delta}^{(m,n)},
\end{align}
we have the four-loop results at NLL~\cite{Caron-Huot:2013fea} and NNLL~\cite{Falcioni:2020lvv} respectively,
\begin{align}
  \label{eq:Delta43}
  \mathbf{\Delta}^{(4,3)} =& 
  - i\pi\frac{\zeta_3}{24}\left[\T_t^2,\left[\T_t^2,\T_{s-u}^2\right]\right]\T_t^2,
  \\
  \label{eq:Delta42}
  \text{Re}\left[\mathbf{\Delta}^{(4,2)}\right] =&
  \,
  \zeta_2\zeta_3,
  \left\{
  \frac14\T_t^2\left[\T_t^2,(\T_{s-u})^2\right]
  +\frac34\left[\T_{s-u}^2,\T_t^2\right]\T_t^2\T_{s-u}^2
  +\frac{d_{AA}}{N_A} - \frac{C_A^4}{24}{}
  \right\},
\end{align}
where only the real part of $\mathbf{\Delta}^{(4,2)}$ is known as we only have NNLL results for the odd amplitude.
By matching the above to an ansatz for the four-loop $\mathbf{\Delta}^{(4)}$ in general kinematics~\cite{Becher:2019avh} we can find the asymptotic limits of the functions~\cite{Falcioni:2021buo}.

The three-loop soft anomalous dimension was shown to be bootstrapable from general considerations and matching to known collinear and high-energy limits~\cite{Almelid:2017qju}. The above results gives useful constraints for a potential four-loop bootstrap.


\section{Conclusions} 
\label{sec:conclusions}

Beyond NLL, in the high-energy limit, amplitudes no longer factorise in terms of a Regge pole, as given in eq.~(\ref{eq:NLLFact}). A non-factorising term, originating from Regge cuts, needs to be included as in eq.~(\ref{eq:NNLL}). This leads to an ambiguity in how to properly define the pole term order-by-order in perturbation theory. In this work, we have shown how to consistently do so with the high-energy properties of Regge poles and Regge cuts in the complex angular momentum plane. We remove all planar MR contributions from the non-factorising term and define a new pole term with new NNLL parameters, namely the two-loop impact factor $\tilde C_i^{(2)}$ and the three-loop Regge trajectory $\tilde\al_g^{(3)}$.

As an application, we find $\mathcal{O}(\eps^2)$ terms of the two-loop impact factors of eqs.~(\ref{eq:gluonDg}) and~(\ref{eq:quarkDq}) using state-of-the-art amplitude computations~\cite{Ahmed:2019qtg}. We also extract, for the first time, the three-loop Regge trajectory, eq.~(\ref{alpha_g_3}), using the amplitudes of ref.~\cite{Caola:2021rqz}. The poles of the Regge trajectory are given by the integrated cusp anomalous dimension when using the cut scheme, eq.~(\ref{eq:cuspSubtractedTraj}).

Furthermore, as a by-product of four-loop computations~\cite{Caron-Huot:2013fea,Falcioni:2020lvv}, we can derive constraints on the structure of infrared divergences in the high-energy limit~\cite{Falcioni:2021buo}, see eqs.~(\ref{eq:Delta43}) and~(\ref{eq:Delta42}).


\printbibliography

\end{document}